\documentclass[%
reprint,
superscriptaddress,
 amsmath,amssymb,
 aps,
 prc,
longbibliography,
dvipsnames,
twocolumn
]{revtex4-1}
\usepackage[main=english,german,ngerman]{babel}

\usepackage{xcolor}
\usepackage{graphicx}
\usepackage{dcolumn}
\usepackage{bm}
\usepackage{siunitx}
\usepackage{dsfont}

\usepackage{mathtools}

\usepackage[shortlabels]{enumitem}

\usepackage{physics}
\usepackage{subcaption}
\usepackage{bbold}
\usepackage{float}
\usepackage{hyphenat}
\newcommand\figref{Figure~\ref}
\usepackage[utf8]{inputenc}
\usepackage{natbib}

\newcommand{\update}[1]{}

\usepackage{hyperref}
\hypersetup{colorlinks=true,citecolor=blue,linkcolor=blue,filecolor=blue,urlcolor=blue,breaklinks=true}

\begin{document}
\newcommand{\orcidicon}[1]{\href{https://orcid.org/#1}{\includegraphics[height=\fontcharht\font`\B]{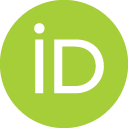}}}

\title{Quantum phase detection generalization from marginal quantum neural network models}
\author{Saverio Monaco \orcidicon{0000-0001-8784-5011}}
\thanks{These authors contributed equally to this work.}
\affiliation{European Organization for Nuclear Research (CERN), Geneva 1211, Switzerland}
\affiliation{Department of Physics, University of Padova, 35122 Padova PD, Italy}

\author{Oriel~Kiss \orcidicon{0000-0001-7461-3342}}
\thanks{These authors contributed equally to this work.}

\affiliation{European Organization for Nuclear Research (CERN), Geneva 1211, Switzerland}
\affiliation{Department of Particle and Nuclear Physics, University of Geneva, Geneva 1211, Switzerland}

\author{Antonio~Mandarino\orcidicon{0000-0003-3745-5204}}
\affiliation{International Centre for Theory of Quantum Technologies, University of Gdańsk, Jana Bażyńskiego 1A, 80-309 Gdańsk, Poland}
 \author{Sofia~Vallecorsa\orcidicon{0000-0002-7003-5765}} 
 \affiliation{European Organization for Nuclear Research (CERN), Geneva 1211, Switzerland}
 
\author{Michele~Grossi\orcidicon{0000-0003-1718-1314}}
\email{michele.grossi@cern.ch}
\affiliation{European Organization for Nuclear Research (CERN), Geneva 1211, Switzerland}

\date{\today}

\begin{abstract}
Quantum machine learning offers a promising advantage in extracting information about quantum
states, e.g. phase diagram. However, access to training labels is a major bottleneck for any
supervised approach, preventing getting insights about new physics. In this Letter, using quantum convolutional neural networks, we overcome this limit by determining the phase diagram of a model where analytical solutions are lacking, by training only on marginal points of the phase diagram, where integrable models are represented.  
More specifically, we consider the axial next\hyp nearest\hyp neighbor Ising (ANNNI) Hamiltonian, which possesses a ferromagnetic, paramagnetic and antiphase, showing that the whole phase diagram can be reproduced.
\end{abstract}

\maketitle

\emph{Introduction.} 
Quantum machine learning (QML) \cite{QML_Lloyd}, where parametrized quantum circuits \cite{PQC} act as statistical models, has attracted much attention recently, with applications to natural sciences \cite{QNN_FF,PRD_anomaly_detection, QML_Higgs,LHC_QML_Higgs, Li_2020, QCL} or in generative modeling \cite{Born_machine_Coyle_2021,QGAN_Zoufal,Zapata_digits,QCBM_Kiss,QCBM_MC}. Even if QML models benefit from high expressivity \cite{abbas_power_2021} and demonstrated superiority over classical models in some specific cases \cite{power_data,covariant_kernel}, it is still unclear what kind of advantage could be obtained with quantum computers \cite{PRXQ_Schuld} in the era of deep neural networks. 

Quantum data, on the other hand, could be a natural paradigm to apply QML, where quantum advantages have already been demonstrated \cite{learning_exp_science}. There is hope that quantum data could be collected via quantum sensors \cite{quantum_sensor}, and eventually directly linked to quantum computers. In this Letter, we emulate the possibility of working with quantum data by constructing them directly on a quantum device. We use a variational ground state solver to obtain approximations of the true ground states in order to mimic noisy real world data. Specifically, this Letter addresses the computation of the phase diagram of the ground states of a Hamiltonian $H$ using a supervised learning approach. Even if similar problems have already been explored for the binary case \cite{Cong_19,PRA_Ising_QPT}, with multiple classes \cite{Multi_QPT_Lazzarin} and computed on a superconducting platform \cite{Wallraff_22_QPD}, all of these approaches suffer from a limitation by construction, a bottleneck. In fact, since labels are needed for the training, and because they are computed analytically or numerically, these techniques can only speed up calculations, but cannot extend beyond their validated domain. Alternatively, anomaly detection (AD), an unsupervised learning technique, has been proposed \cite{PRL_Kottman,PRR_anomaly_detection} as a way to bypass this bottleneck, by finding structure inside the data set. However, AD can only obtain qualitative, possibly unstable, results and the performance can therefore be difficult to assess. Instead, the proposed approach provides a clear prediction for the boundaries of the adopted model, with the possibility to evaluate the performance on a validation set.

This Letter numerically demonstrates that QML can make predictions to regions where analytical labels do not exist, after being only trained on easily computable subregions. Moreover, QML only needs very few training labels to do so, as already pointed out by \textcite{generalisation_few_data, Banchi}. In particular, we make a step toward an out\hyp of\hyp distribution generalization \cite{Caro_out_of_distribution}, where the training and testing set do not belong to the same data distribution, which is known to be a difficult task \cite{out_hard}. This drastically changes the perspective, extending QML capabilities to extrapolate and eventually discover new physics when trained on well\hyp established simpler models. \\

\noindent \emph{The model.}
We consider the axial next\hyp nearest\hyp neighbour Ising (ANNNI) model 

\begin{equation}
\label{ANNNI}
    H =   J \sum_{i=1}^{N} \sigma_x^i\sigma_x^{i+1} - \kappa \sigma_x^{i}\sigma_x^{i+2} + h \sigma_z^i,
\end{equation}
where $\sigma^i_a$ are the Pauli matrices acting on the $i$th spin, $a\in \{x, y, z\},$ and we assume open boundary conditions. The energy scale of the Hamiltonian is given by the coupling constant $J$ (without loss of generality we set $J=1$), while the dimensionless parameters $\kappa$ and $h$ account for the next-nearest-neighbor interaction and the transverse magnetic field, respectively.  We restrict ourselves to $\kappa\geq 0$, $h\geq 0$ and even $N$.
The difference of sign between the nearest and next-nearest interactions, leading to a ferro- or antiferro-magnetic exchange in the system, is
responsible for the magnetic frustration. Thence, the ANNNI model offers the possibility to study the competing mechanism of quantum fluctuations 
due to the transverse magnetic field and frustration. The phase diagram of the quantum model at $T=0$K temperature has been studied mainly by renormalization group or Monte Carlo techniques in $d$ dimensions exploiting also the correspondence with the classical analog in $d+1$ dimensions \cite{ANNNI_report, PRB_ANNNI, PRE_ANNNI, PRB_DMRG_ANNNI, PRB_floating_phase, ANNNI_19}.
The phase diagram is quite rich and three phases have been confirmed, separated by two second\hyp order phase transitions. The first, for low frustration $(\kappa<0.5)$ of the Ising type separates the ferromagnetic and the paramagnetic phases along the line $h_I(\kappa) \approx \frac{1 - \kappa}{\kappa} \left(1 - \sqrt{\frac{1 - 3 \kappa + 4 \kappa^2 }{1 - \kappa}} \right)$. The other one of a commensurate-incommensurate type appears between the paramagnetic phase and an antiphase for values of the field  $h_{C}(\kappa) \approx 1.05 \sqrt{(\kappa - 0.5) (\kappa - 0.1)}$, in the high frustration sector $(\kappa>0.5)$. As usual, the paramagnetic phase is the disordered one, in contrast with the two ordered phases: the ferromagnetic and the antiphase one. In particular, they are different because the former is characterized by all the spins aligned along the field direction, and the latter has a four-spin periodicity, composed of repetitions of two pairs of spins pointing in opposite directions. The point $\kappa=0.5$ represents a multicritical point.

We mention here that other relevant lines have been numerically addressed but not confirmed. One signaling an infinite-order phase transition of the Berezinskii–Kosterlitz–Thouless (BKT) type for $h_{BKT}(\kappa) \approx  1.05(\kappa - 0.5),$ delimiting a floating phase between the paramagnetic and the antiphase \cite{PRB_floating_phase}, and a disorder line where the model is exactly solvable known as the Peschel-Emery (PE) line $h_{PE}(\kappa) \approx \frac{1}{4 \kappa} - \kappa $ \cite{peschel1981,PRB_DMRG_ANNNI}. \\
\\
\noindent \emph{Variational state preparation.}
The purpose of the variational quantum eigensolver (VQE) \cite{vqe_original} is to calculate the ground state energy
of a Hamiltonian $H(\kappa,h)$ on a quantum computer. Using the Rayleigh\hyp Ritz variational principle, the VQE minimizes the energy expectation value of a parametrized wavefunction and has been successfully applied in quantum chemistry \cite{VQE_Gambetta,Panos_excitation_preserving,UCC-chemistry}, in nuclear physics \cite{Papenbrock-Deuterium, Be8-VQE, VQE_Li6} or in frustrated magnetic systems \cite{VQE_magnet,Grossi_LMG_VQE}.
Here, we are interested in the final eigenstates, represented by an ansatz $\ket{\psi(\theta;\kappa,h)}$, to be used as quantum data.
 Typically, the ansatz is chosen as a hardware\hyp  efficient (HEA) quantum circuit \cite{VQE_Gambetta, ansatz}, which is built with low connectivity and gates that can be easily run on noisy intermediate\hyp scale quantum (NISQ) \cite{Preskill2018quantumcomputingin} devices. For instance, we use $D=6 (9)$ repetitions of a layer consisting of free rotations around the $y$ axis $R_y(\theta)=e^{-i\theta \sigma_y/2}$ and controlled-NOT (CNOT) gates with linear connectivity CX$_{i,i+1}$ for $0\leq i<N$ \cite{nielsen_chuang_2010}, for $N=6(12)$ spin systems. The optimization is performed using the gradient\hyp descent\hyp based ADAM algorithm \cite{adam}, with an initial learning rate of 0.3 and a parameter recycling scheme to improve the convergence \cite{param_recycling}. Moreover, we note that the VQE can also be used to recursively compute excited states \cite{Excited-states}, which we used to show that the ground states of the ANNNI model are only degenerate at the boundaries in the phase diagram, where the ground states corresponding to the different phases are competing, excluding the bit flip symmetry at $h=0$. Finally, we asses the accuracy of the VQE states by comparing with the exact energy and observe that the relative error ratio is always below 1\%. Moreover, it seems that the energy accuracy distribution is able to reveal the Peschel\hyp Emery line, since the predicted energy values are more accurate along it. More details about the implementation, optimization, degeneracy and accuracy can be found in Appendix \ref{app:vqe}.
\\
\\
\noindent \emph{Quantum convolutional neural networks (QCNNs).}
QCNNs are a class of quantum circuits, inspired by classical convolutional neural networks (CNN) \cite{lecun}, originally proposed in \cite{Cong_19}. The QCNN is trained to detect quantum phase transitions, effectively learning an observable $O(\theta)$ that linearly separates two states $\ket{\psi_A}$ and $\ket{\psi_B}$ from two different phases $A$ and $B$, such that $\expval{O(\theta)}{\psi_A}<0<\expval{O(\theta)}{\psi_B}$ \cite{science_robert_ml}, which exist since the phases in the ANNNI model are not topological. Intuitively, non topological phases of matter exhibit macroscopic differences, which can be captured by the variational observable $O(\theta)$. In principle, quantum phase detection could be performed by measuring different string order parameters (SOPs) \cite{Cong_19}. However, the SOP vanishes near the phase transition, thus requiring exponentially many samples for the classification. On the other hand, the QCNN output is much  sharper, therefore reducing the sample complexity. This changes quantum phase detection to the task of designing and training an appropriate ansatz.

\begin{figure}
    \centering
    \includegraphics[scale=0.22]{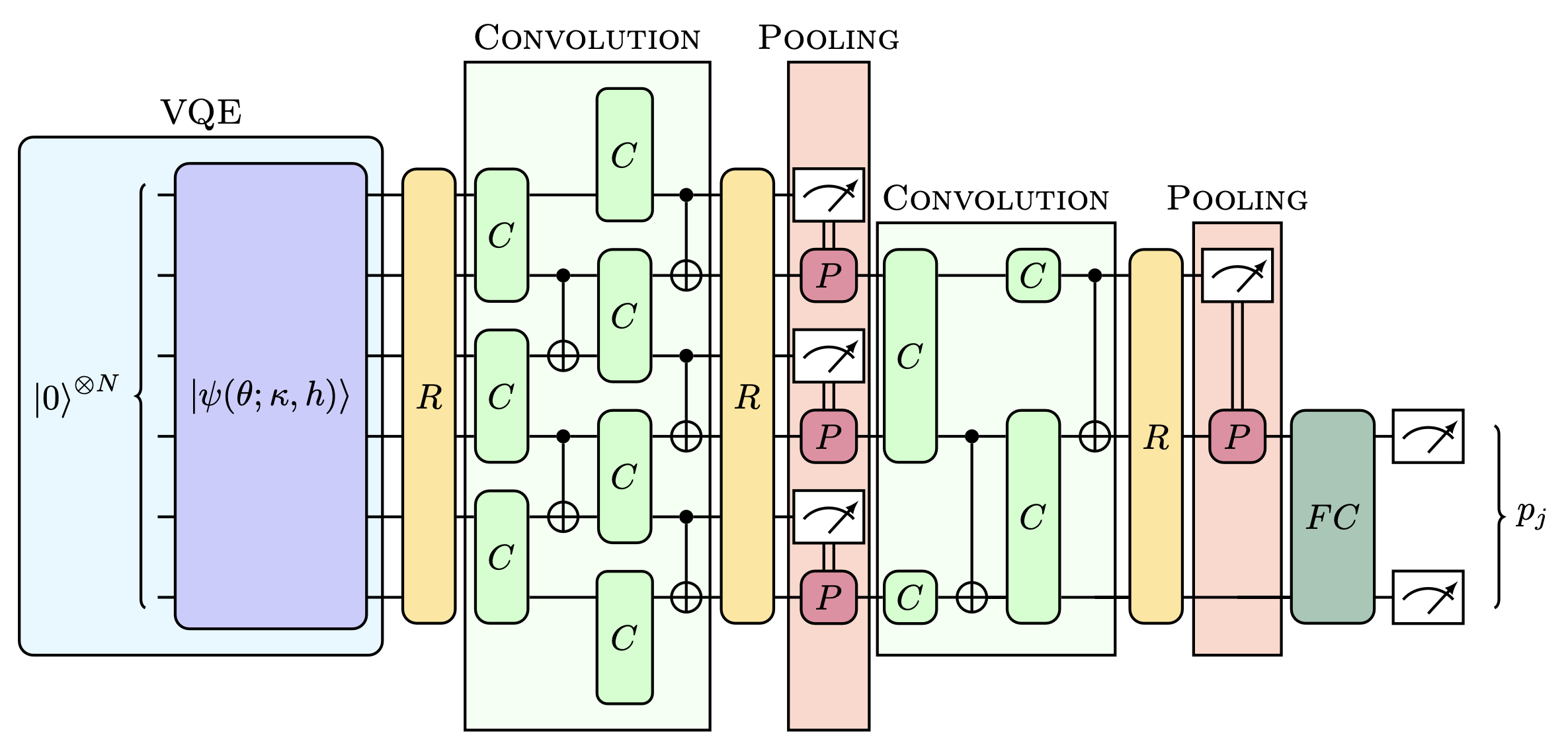}
    \caption{Circuit architecture: VQE states (blue) are the input of the quantum convolutional neural network composed of free rotations $R$ (yellow), convolutions $C$ (light green), pooling $P$ (red), and a fully connected layer $F$ (dark green).}
    \label{fig:qcnn}
\end{figure}

In our implementation, the QCNN starts with a free rotation layer around the $y$ axis, followed by blocks consisting of convolutions, free rotations, and pooling layers that halve the number of qubits to $k$ until $k= \lceil \log_2{(K)}\rceil$, where $K$ it the total number of quantum phases. Finally, a fully connected layer and measurement are performed in the computational basis.
An example with $N=6$ qubits is shown in \figref{fig:qcnn} where we have free $y$ axis rotations (yellow), $R(\vec{\theta}) = \bigotimes_{i=1}^{N} R_y(\vec{\theta}_i)$, two\hyp qubit convolutions (light green) $C(\theta) = \bigotimes_{i=1}^{2} R_y(\theta)$, pooling (red) $P(\vec{\theta},\phi,b) = R_y(\vec{\theta}_b)R_x(\phi)$ with $b\in \{0,1\}$ the value of the measured qubit, and a two\hyp qubit fully connected (dark green) gate\ $ F(\vec{\theta}^{(1)},\vec{\theta}^{(2)})=\left(\bigotimes_{i=1}^{2} R_y(\vec{\theta}^{(i)}_1)R_x(\vec{\theta}^{(i)}_2)R_y(\vec{\theta}^{(i)}_3)\right)\text{CX}_{1,2}$. \\
 
QCNNs have been shown to be resistant to barren plateaus \cite{BP_QCNN} due to their distance from low $T2$ design and are therefore good candidates for any quantum learning tasks. The analogy with CNN holds in the quantum settings since convolution and pooling layers are functions of shared parameters and the reduction of the circuit's dimension is guaranteed by the intermediate measurement. The whole algorithm flow starts with the QCNN taking as input ground states $\ket{\psi(\theta;\kappa,h)}$ from the Hamiltonian family $H(\kappa,h)$, obtained through the VQE. The quantum network then outputs the probability $p_j(\kappa,h$) of being in one of the $K=3$ phases (ferromagnetic, paramagnetic or antiphase), where $p_j(\kappa,h)$ is computed as the probability of measuring the state $\ket{01},\ket{10},\ket{11}$  on the two output qubits. Since the phase diagram of the ANNNI model only contains three phases, the state $\ket{00}$ is interpreted as a \emph{garbage} class.\\ 

\noindent \emph{Generalization.} 

\begin{figure}
    \centering
    \includegraphics[scale=0.35]{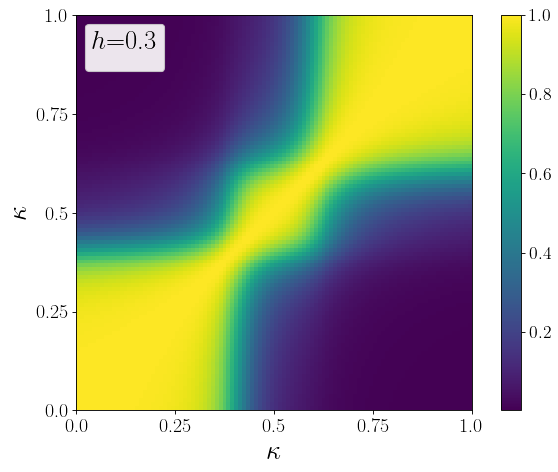}
    \caption{Fidelity between the ground states of the ANNNI model at $h=0.3$ and $N=12$. We observe three different cluster, corresponding to the ferromagnetic, paramagnetic, and antiphase, respectively.}
    \label{fig:tile_like}
\end{figure}

The main contribution of this Letter is to demonstrate the ability of QCNN to work in a partial supervised approach and thus get closer to an out\hyp of\hyp distribution generalization by training on a set of easily available labels. We first argue that this generalization is expected to hold according to \cite{Banchi} if the ground states of the ANNNI model are clustered, i.e., if the fidelity between states in the same phase is high \cite{fid1, fid2,fid3}, while being low between different phases. This is indeed the case as shown in Figure \ref{fig:tile_like} along the line $h=0.3$ for the $N=12$ spin case.
Even if the requirements of the generalisation results from \cite{generalisation_few_data} do not hold since the training data are only located on the boundaries, and specifically not independent and identically distributed (i.i.d.), we observe a numerical agreement with the generalization error's scaling behavior predicted in Ref.~\cite{generalisation_few_data}, i.e.,  $\mathcal{O}\left(\sqrt{\frac{T}{n}}\right)$, where $T$ is the number of parameters and $n$ the number of training points. Since the QCNN is composed of $T=\mathcal{O}[\log{(N)}]$ parameters \cite{Cong_19}, we can control the expected risk by training on $n = \mathcal{O}\left[\log{(N)}\right]$ points. \\ 

\noindent \emph{Training set.} The training data set consists of the composition of points from two analytical models derived from the simplification of the physical model used. Specifically, we consider the integrable Ising model in transverse field in the case $\kappa=0$ and the \emph{quasi classical} model
when $h=0$, where quantum fluctuations no longer exist. We demonstrate that QCNNs extend their prediction to the all phase diagram when only trained on the marginal model given by $\mathcal{S}_X^n \subseteq \{(\kappa,h) \in \{0\} \times [0,2] \} \cup \{(\kappa,h) \in  [0,1] \times \{0\} \}$. We consider three types of subsets $X\in \{GC,G2, U\}$, $\mathcal{S}_{GC}^n$ where $n$ training points are sampled normally around each critical point \{(0,1), (0.5,0)\}, $\mathcal{S}_{G2}^n$ where $n$ training points are sampled normally at the middle of each phase \{(0,1.5), (0,0.5), (0.25,0), (0.75,0)\} and $\mathcal{S}_U^n$ where $n$ data points are drawn uniformly on both axes. The QCNN is trained using the cross entropy $\mathcal{L}$ loss,
\begin{equation}
   \mathcal{L} =  -\frac{1}{|\mathcal{S}_X^n|}\sum_{(\kappa,h)\in \mathcal{S}_X^n} \sum_{j=1}^{K} y_j(\kappa,h) \log{(p_j(\kappa,h))}
\end{equation}
between the one\hyp hot classical labels $y_j(\kappa,h)$ and the predictions on the training region $\mathcal{S}_X^n$ of the phase space.\\

\noindent \emph{Results.}
\begin{figure*}
\centering
\centering
 \subcaptionbox*{}{\includegraphics[width = 2.1in]{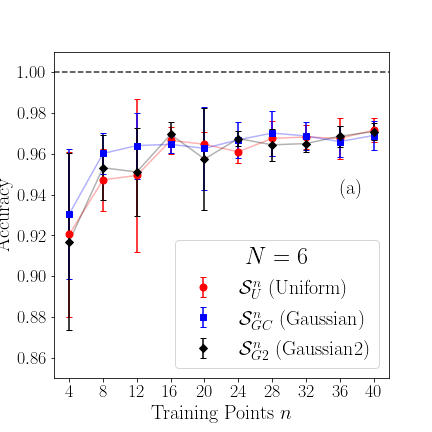} }\hskip -0ex
 \subcaptionbox*{}{\includegraphics[width = 1.9in]{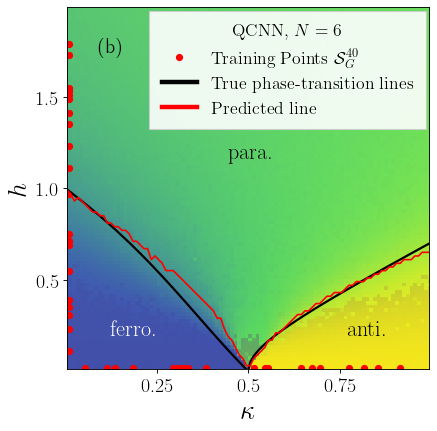} }\hskip -0ex
 \subcaptionbox*{}{\includegraphics[width=2.75in]{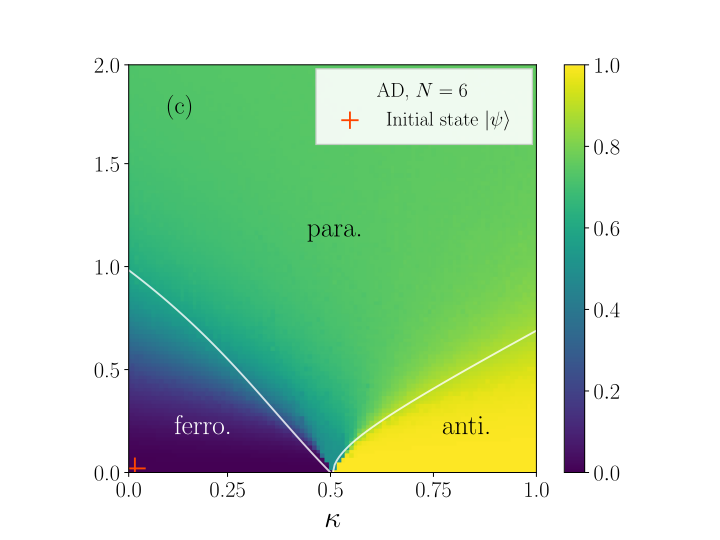} }
 \vskip -8ex
\centering
 \subcaptionbox*{}{\includegraphics[width=2.1in]{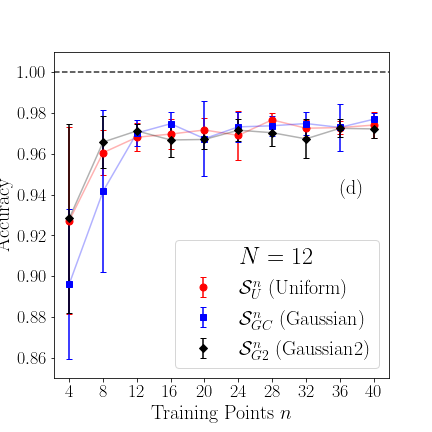} }\hskip -0ex
  \subcaptionbox*{}{\includegraphics[width=1.9in]{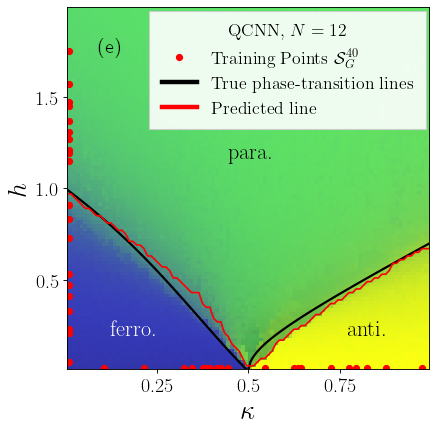} }\hskip -0ex
  \subcaptionbox*{}{\includegraphics[width=2.75in]{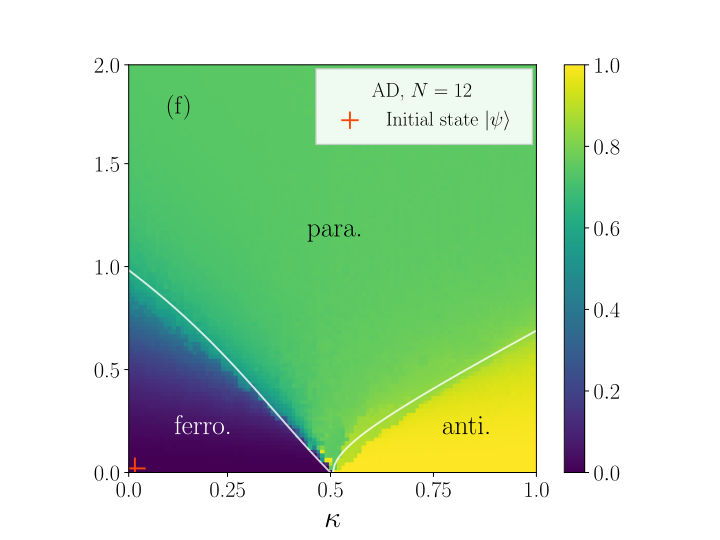} }
 \vskip -4ex
\caption{Quantum phases classification. (a) shows the classification accuracy of the QCNN as a function of the number of training points $n$, for the Gaussian centered around the critical points ($GC$ blue), Gaussian centered around the middle of each phase ($G2$ black) and uniform sampling (red). (b) displays the phase diagram predicted by the QCNN trained on $\mathcal{S}_{G2}^{40}$ (red dots) where the color represents the probability mixture of being in one of the three phases and the red lines the predicted boundaries, while (c) shows the anomaly score for a $N=6$ spins systems trained on the initial state $\ket{\psi}$ (red cross). (d)-(f) are similar but for $N=12$ spins. The black lines are $h_I(\kappa)$ for $\kappa<0.5$ and $h_C(\kappa)$ for $\kappa>0.5$}
\label{fig:phase_diag}
\end{figure*}

Once we have introduced the problem and defined the techniques used, we can analyze the quality of the results obtained under ideal conditions with a quantum simulator. 

We study our ability to reconstruct the phase diagram of the ANNNI model, characterized by a non trivial disordered paramagnetic phase, the ordered ferromagnetic phase, and antiphase one. To test the stability of the proposed approach, we consider the model with an increasing number of spins $N=6,12$ and sampling a different number of points $0<n\leq 100$ used for the training. By virtue of the quality of the results, we evaluated the influence of different sampling of the training points corresponding to the two physical models that could affect the quality of the classification. A summary of the results can be qualitatively seen in \figref{fig:phase_diag}, while more quantitative results for the QCNN are displayed in Appendix \ref{app:pi}. In the first row, we have the phase diagram reconstruction for the ANNNI model with six spins, where the black lines represent the analytical transition explained above in the model section. The second line in the figure shows the same for a system with $N=12$ spins.

The first column shows the accuracy, computed on the whole phase space, as a function of the number of training points $n$, for the Gaussian centered around the critical points $X=GC$ (blue), around the middle of each phase $X=G2$ (black), and the uniform $X=U$ (red) sampling scheme, where the error bars correspond to one standard deviation from ten independent runs. We observe that the accuracy quickly increases with $n$, before saturating for $n\geq 20$, as argued in Ref. \cite{generalisation_few_data} and that the sampling strategy does not play a major role. More importantly, sampling away from the critical points is enough. The second column displays the phase diagram obtained with the QCNN trained on $n=40$ points. Colour shades represent the continuous probability distribution of the QCNN for our multiclass classifier as a probability mixture (blue, green and yellow times the relevant probability), while the red lines represent the predicted boundaries. The individual probabilities of each phases predicted by the QCNN are shown in Appendix \ref{app:pi}. The last column instead shows the comparison to the unsupervised learning approach inspired from \cite{PRR_anomaly_detection} where the autoencoder is trained to compress the single red cross $\ket{\psi}$, and tested on the remaining points. In a nutshell, the autoencoder is expected to perform poorly if the states are far away in the Hilbert space, i.e., if they belong to different phases, thus leading to a high compression score. The color scale shows the compression loss of each state. Additional details about the implementation of the anomaly detection can be found in Appendix \ref{app:AD}. It is worth noting that although only one training point is sufficient to obtain a qualitatively good phase diagram, only QCNNs allow a quantitative prediction for the phase. Moreover, while the QCNNs are stable when changing the training set, it is easy to find initial states where AD performs poorly, for instance by starting in the paramagnetic phase. The relatively good performance of AD can be explained by the product state nature of the training point. Hence, the product state can be easily compressed with the autoencoder \cite{Q_encoder_Romero}, while states corresponding to a high magnetic field cannot.

\noindent \emph{Conclusions.}
This Letter addresses the computation of the phase diagram of a non integrable model, by training a QCNN on the limiting integrable regions of the considered ANNNI model. We provide numerical evidence that the QCNNs are able to generalize from non\hyp i.i.d. training data, which is a challenging task in general. The numerical simulations suggest that QCNNs can carry this task with more than $97\%$ accuracy, using only $n=20$ quantum data points distributed on the two integrable axes of the phase space. Moreover, the data points do not need to be close to the critical points. The accuracy of the QCNN quickly increases to reach its maximum as a function of the number of training points, suggesting that QCNNs can generalize from a few data points. Being a supervised method, the QCNN is not able to detect phases that are not present in the training set $\mathcal{S}_X^n$, i.e., the boundaries, such as the BKT phase transition and the PE line. Nevertheless, AD is also not able to reveal them and is limited to qualitative predictions, while a supervised approach gives quantitative results whose quality can be easily evaluated on the validation set. Moreover, by approaching out\hyp of\hyp distribution generalization, we propose a solution to the bottleneck of needing training labels, that are generally challenging to obtain. Consequently, we make a step into extending the reach of QML to useful applications in physics. Future work should be performed to detect phases not present in the training set, such as the floating phase or the PE line, by either affording $\mathcal{O}(1)$ training points inside these unrepresented phases or mixing the QCNN with the unsupervised approach. \\ 
\\

\section*{Acknowledgments} The authors would like to thank Zo\"e Holmes for useful discussions about generalization. S.M., O.K., and M.G. are supported by CERN through the CERN Quantum Technology Initiative. A.M. is supported by Foundation for Polish Science (FNP), IRAP project ICTQT, contract No. 2018/MAB/5, co financed by EU Smart Growth Operational Program, and (Polish) National Science Center (NCN), MINIATURA DEC-2020/04/X/ST2/01794.

\section*{Code Availability}

The code used to generate the data set and the figures of the present Letter is publicly available \cite{software}.

\bibliography{bibliography} 

\appendix 

\section{Quantum data set} 
\label{app:vqe} 
\begin{figure}
    \centering
    \includegraphics[width=2.1in]{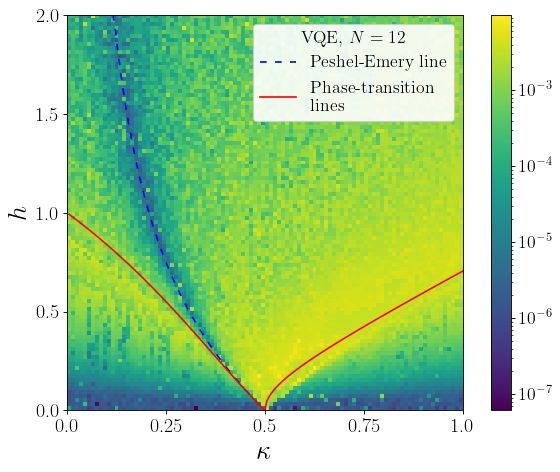}
    \caption{Normalised difference between the energy obtained with exact diagonalization and those obtained with the VQE on a log scale. Red lines correspond to the numerical reference boundaries obtained through Monte Carlo, while the dashed blue line represents the second\hyp order PE line.}
    \label{fig:acc12}
\end{figure}

\begin{figure}
    \centering
    \includegraphics[width=2.1in]{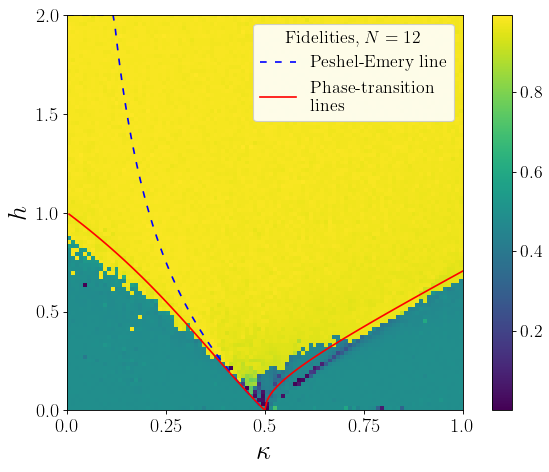}
    \caption{Fidelity between the VQE and exact ground states for $N=12$ spins. Red lines correspond to the numerical reference boundaries obtained through Monte Carlo, while the dashed blue line represents the second\hyp order PE line.}
    \label{fig:fid12}
\end{figure}

The variational quantum eigensolver (VQE) is used to construct a quantum data set in the form of quantum circuits representing the ground states of the ANNNI model. In a nutshell, a parameterized wavefunction is prepared on a quantum computer, and its parameters are iteratively updated to minimize the expected value of the energy. Hence, using the variational principle, this parameterized wavefunction shall be a good approximation of the ground state, assuming that the ansatz is expressive enough and that the optimization is successful. We recall that the ansatz is constructed with $D=6 (9)$ repetitions of a layer consisting of free rotations around the $y$\hyp axis $R_y(\theta)=e^{-i\theta \sigma_y/2}$ and CNOT gates with linear connectivity CX$_{i,i+1}$ for $0\leq i<N$, for $N=6(12)$ spins. The optimization is performed with the ADAM algorithm, which uses first\hyp order gradient descent with momentum, where the gradients are obtained with the automatic differentiation framework provided by Pennylane \cite{pennylane}. We note that the gradients can be obtained on quantum hardware using the parameter\hyp shift rule \cite{quantum_grad}. To improve the convergence and accuracy of the quantum data set, we perform five optimization rounds composed of 1000 update steps by reducing the initial learning rate from 0.3 to 0.1 between each round. Moreover, we use a parameters recycling scheme \cite{param_recycling}, where the initial parameters at the point ($h,\kappa$) are chosen to be the converged parameters from the previously computed neighboring site. This strategy improves the speed and accuracy of the optimization while keeping a high fidelity between neighboring states, as expected from a physical point of view. Figure \ref{fig:acc12} shows the accuracy of the VQE ground states energy with respect to the exact one for the $N=12$ spins cases. We observe that the relative error ratio
\begin{equation}
    \Delta E = \left|\frac{E_{\text{VQE}}-E_{\text{exact}}}{E_{\text{exact}}} \right|
\end{equation}
is always below 1\%. Moreover, we notice that the accuracy is significantly better on the Peschel\hyp Emery disorder line, indicating that the accuracy of the VQE algorithm could be used to detect interesting regions of the phase diagram. For instance, it could be used in conjunction with the DMRG algorithm to expose such disorder lines. 
We also report the fidelity between the VQE and exact ground states in Figure \ref{fig:fid12}. We observe that the paramagnetic states are almost exactly reproduced, while the fidelity for low magnetic field values is about 50\%, and lower around the critical points. 

\section{Details on the predicted phase diagram}
\label{app:pi}
\begin{figure}
\centering
\centering
 \subcaptionbox{}{\includegraphics[width=2.1in]{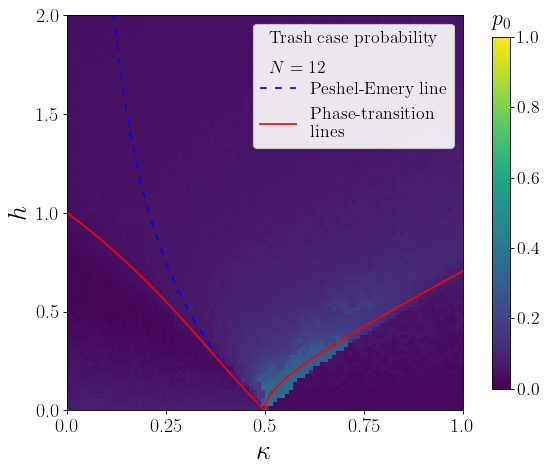} }\hskip +1ex 
 \subcaptionbox{}{\includegraphics[width=2.1in]{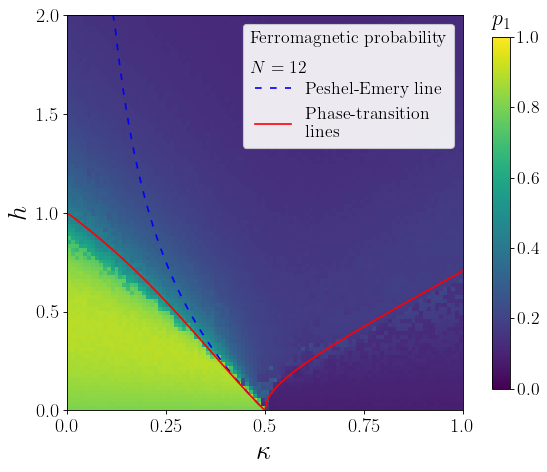} }\hskip -14ex

 \vskip -1ex
\centering
 \subcaptionbox{}{\includegraphics[width=2.1in]{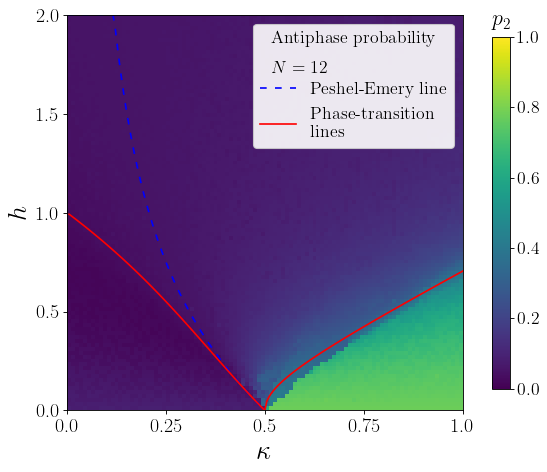} }\hskip +1ex
  \subcaptionbox{}{\includegraphics[width=2.1in]{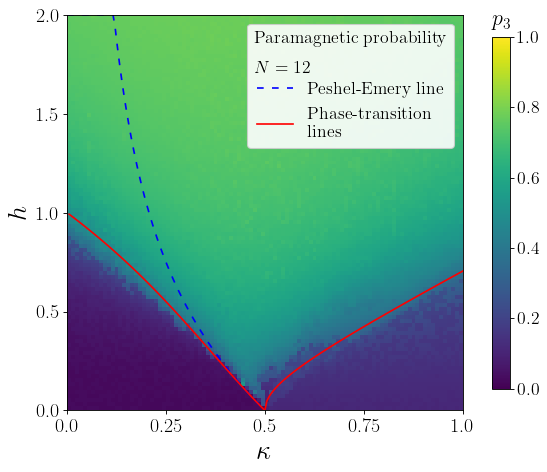} }\hskip -14ex
 
 \vskip 0ex
\caption{Individual probability $p_i(\kappa,h)$ , $0\leq i <4$ (a-d), as a function of the Hamiltonian parameters. Red lines correspond to the numerical reference boundaries obtained through Monte Carlo, while the dashed blue line represents the second\hyp order PE line.}
\label{fig:pi}
\end{figure}

As explained in the result section, the QCNN outputs the probability $p_j(\kappa,h)$ of obtaining the four two\hyp qubit state $\ket{01}, \ket{10}, \ket{11}$ and  $\ket{00}$. We associate the first three to the physical ferro\hyp, para\hyp magnetic, and antiphase, while the last one is treated as a garbage class. In the main text, we plot the probability mixture of the three physical phases with a blue\hyp green\hyp yellow color channel. For clarity and completeness, the individual probability $p_i(\kappa,h)$ with $N=12$ spins are shown in Figure \ref{fig:pi}, where the color indicates the magnitude of the corresponding probability. 

We observe that the probability of being in the garbage class is almost zero, except near the triple point. Moreover, all predicted phase boundaries are sharp, indicating the classifier's confidence. 

\section{Anomaly Detection}
\label{app:AD}
For the reader's convenience, we will recall the unsupervised anomaly detection (AD) scheme, initially proposed by \textcite{PRR_anomaly_detection}, to draw the phase diagram of the Bose\hyp Hubbard model. Since it is an unsupervised learning technique, it bypasses the bottleneck of needing classical training labels and is, therefore, an alternative to the approach taken in this letter. 

\begin{figure}
    \centering
    \includegraphics[scale=0.28]{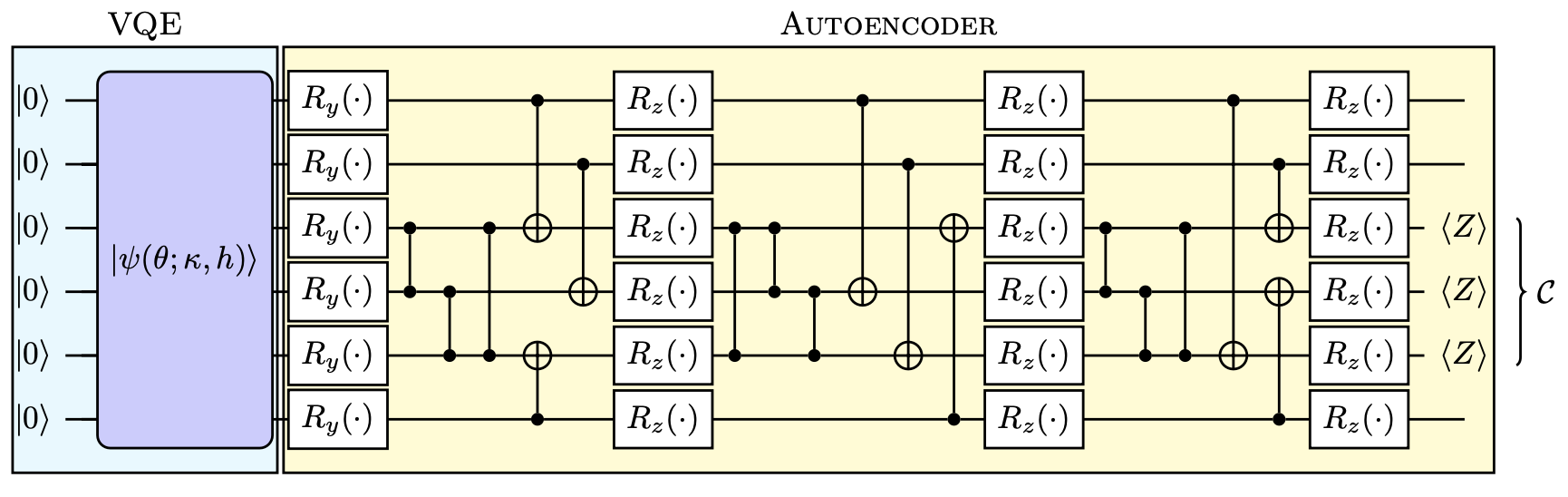}
    \caption{Compression circuit (yellow) and anomaly score measurement ($\mathcal{C}$) of the ground states of $H(\kappa,h)$ obtained through a VQE (blue). The $\cdot$ represents independent parameters.}
    \label{as}
\end{figure}

As a first step, an initial state $\ket{\psi}$ is chosen in the data set composed of the ground states of $H$. Although there is no formal restriction, it should lie far from any critical points. A quantum encoder \cite{Q_encoder_Romero} is then trained to learn to compress $\ket{\psi}$ on a $k$\hyp qubit state $\ket{\phi}$, with quantum register $q_C$ and $k<N$, i.e., to write $\ket{\psi} = \ket{\phi}\otimes \ket{T}$, where the latter is a $(N-k)$\hyp qubit trash state with register $q_T$. The registers here refer to the indices of the qubits composing the states. In practice, an anomaly score based on the Hamming distance between the trash state $\ket{T}$ to $\ket{0}^{\otimes (N-k)}$, written as 
\begin{equation} 
\mathcal{C} = \frac{1}{2}\sum_{j\in q_T} (1-\left<Z_j\right>),
\end{equation}
and we make the choice $k=\lfloor N/2 \rfloor$.
Intuitively, the encoder compresses similar states, i.e., states in the same phase, with success but will fail to compress states in a different phase, leading to a high anomaly score. 
The encoder, as proposed in \cite{PRR_anomaly_detection}, is composed of $D$ layers of independent $R_y(\theta)$ rotations on all qubits and CZ$_{i,j}$ gates for $i\in q_C,j\, \in q_T$ and  $i,j \in q_T$. We use a slightly modified version, with a first layer of $R_y(\cdot)$ individual rotations, followed by $D=3$ layers composed of  CX$_{i,j}$ gates for $i \in q_C$ and $j\in q_T$, CZ$_{i,j}$ gates with $i,j \in q_T$ and independent $R_z(\cdot)$ rotations as displayed in \figref{as} for $N=6$.

We highlight a few differences with the supervised approach. First, the anomaly score measurement is highly dependent on the choice of the initial state $\ket{\psi}$, and can often lead to phase diagrams without any clear phase separation. Moreover, there is no quantitative way to assess the validity of the predicted phase diagram, while with the QCNN, we may evaluate the accuracy on the validation set. Finally, the anomaly score only provided qualitative results. Hence, only a continuous number (the anomaly syndrome) is associated with each point, and there is no canonical way to assign it to a particular phase. On the other hand, the QCNN outputs the probability of being in each phase and therefore, a unique predicted phase can be assigned to each quantum state.

\end{document}